\documentclass[12pt,twocolumn]{article}

\usepackage{hyperref} 
\hypersetup{
    bookmarks=true,         
    unicode=false,          
    pdftoolbar=true,        
    pdfmenubar=true,        
    pdffitwindow=false,     
    pdfstartview={FitH},    
    pdftitle={Radio Pulsar Navigation},      
    pdfauthor={Dong Jiang},                  
    pdfsubject={Radio Celestial Navigation}, 
    pdfcreator={DJ},                         
    pdfproducer={DJ},                        
    pdfkeywords={Radio Pulsar Navigation},   
    pdfnewwindow=true,      
    colorlinks=true,        
    linkcolor=red,          
    citecolor=green,        
    filecolor=magenta,      
    urlcolor=cyan           
}

\usepackage{natbib}
\usepackage{graphicx}
\usepackage{aas_macros}

\begin{document}

\title{Pulsar Navigation in the Solar System}

\author{Jiang Dong,
\thanks{DJ is in YunNan Astronomical Observatory, 
E-mail: \href{mailto:djcosmic@gmail.com}{djcosmic@gmail.com}, 
and Scientific and application center of lunar and deep space exploration, NAOs, China. The content of this paper have applied the patent.}
\thanks{Manuscript received ---; revised ---.}}

\maketitle

\begin{abstract}
The X-ray Navigation and Autonomous position Verification (XNAV) is tested
which use the Crab pulsar 
under the Space Test Program that use starlight refraction.
It provide the way that the spacecraft could autonomously
determine its position with respect to an inertial origin.
Now we analysis the sensitivity of the exist instrument and the signal process
that use radio pulsar navigation
and discuss the integrated navigation which use radio pulsar,
then give the different navigation mission analysis and design process basically
which include the space, the airborne, the ship and 
the land of the planet or the lunar. Our analysis show that 
we will have the stability profile (signal-to-noise is 5 ) that 
use a 2 meters antenna observe some strong sources of radio pulsar
in 36 minutes which based on the today's technology. 
So the pulsar navigation can give the continuous position in deep space, 
that means we can freedom fly successfully in the solar system use
celestial navigation that include pulsar and traditional star sensor.
It also can less or abolish the dependence to Global Navigation
Satellite System (GNSS) which include GPS, GRONSS, Galileo and BeiDou et al. 
\end{abstract}

\section{Introduction}
The gigantic success of celestial navigation is  
Christopher Columbus awareness of the American continents 
in the Western Hemisphere in 1492,
even if it first be discovered in 1421 by ZhengHe or Viking, 
they all use celestial navigation in the road. 
The history of navigation use celestial origin from the ancient activity 
which include hunting and back to home in the night. 
The base principle of celestial navigation is used the moon, stars, 
and planets as celestial guides assuming the sky was clear in that times. 

With the period of the Age of Discovery or Age of Exploration 
(the early 15th century and continuing into the early 17th century), 
many new technique be invented which include water clock, 
quadrant Sextant (1731, John Hadley), Chronometer (1761, John Harrison),
the Sumner Line or line of position (1837, Thomas Hubbard Sumner),
the Intercept Method or Marcq St Hilaire method (1875, Marcq St Hilaire) et al.
The Sumner Line and Marcq St Hilaire method construct
the foundation of modern nautical navigation. 
The lighthouse are used to mark dangerous coastlines, hazardous shoals 
and reefs, safe entries to harbors and can also assist in aerial navigation. 
For the physics and chemistry development,
Gustaf Dal\'{e}n in recognition of his remarkable invention of 
automatic valves designed to be used in combination with 
gas accumulators in lighthouses and light-buoys (the noble prize in 1912).

After Guglielmo Marconi achieve the radio communication in 1895,
navigation that use radio as an aid has been practiced in Germany since 1907. 
Scheller invent the complimentary dot-dash guiding path, 
which can be seen as a `landmark' for several decades of navigational aids. 
The first practical VHF radar system be installed on French ship in 1935.
Radio navigation grow fast for the defense technique need 
in the World War II, the LOng RAnge Navigation (Loran) system is invented.
In 19 century, it is proposed that use artificial stellar navigation.
Until 1960, the first satellite navigation system, Transit,
is first successfully tested, used by the United States Navy.
Then it evolve to Global Positioning System (GPS).
The soviet also build the similar system - GLONASS for the same reason.
Now the Global Navigation Satellite system (GNSS) in the realism or dreams
still have BeiDou, GALILEO et al.

An Inertial Navigation System (INS) is a navigation aid that 
uses the computing and motion sensors to continuously track the position, 
orientation, and velocity (direction and speed of movement) of a moving object 
without the need for external references. 
The gyroscopic compass (or gyro compass) was introduced in 1907.
In 1942, the first INS be applied in V2 missile,
then it be used in aeronautics, nautical and space widely.

When GPS and INS is still not ripe,
Celestial Navigation System (CNS) be spread to aeronautics by
US (B-52, B-1B, B-2A, C-141A, SR-71, F22 et al.) and
Soviet (TU-16, TU-95, TU-160 et al.) \citep{pdl+01, w07}.
Then the star tracker 
(i.e. track one star or planet or angle between it) \citep{mf63} 
be used to determine the attitude of the spacecraft
in help orient the Apollo spacecraft enroute to and from the Moon.  
Now the advanced star sensor (i.e. sense many star simultaneous) 
has been developed for the application of optical CCD technique  \citep{khh92}.

Although GPS and INS almost can finish any job in this planet now,
someone still think celestial navigation is important for 
it can be used independently of ground aids and has global coverage, 
it cannot be jammed (except by clouds) and 
does not give off any signals that could be detected by an the others.
The traditional maritime state which include US, Russia, UK and French,
all spend many money in CNS for its unique advantage.

\section{X-ray Pulsar-based  Navigation  System in Spacecraft}
After radio pulsar be discovered by Bell, J. and Hewish, A. in 1967,
Downs, G. S. give the advice that use radio pulsars for 
interplanetary navigation in 1974 \citep{d74}. 
But in that time, both the radio and optical signatures from pulsars 
have limitations that reduce their effectiveness for spacecraft navigation. 
At the radio frequencies that pulsars emit (from 100 MHz to a few GHz)
and with their faint emissions, 
radio-based systems would require large antennas 
(on the order of 25 M in diameter or larger) to detect sources, 
which would be impractical for most spacecraft. 
Also, neighboring celestial objects including 
the sun, moon, Jupiter, and close stars, 
as well as distance objects such as radio galaxies, quasars,
and the galactic diffuse emissions, are broadband radio sources that
could obscure weak pulsar signals \citep{rwp06, s05, spr+06}.
So Chester, T. J. and Butman, S. A. describe spacecraft navigation 
using X-ray pulsars in 1981 \citep{cb81}.
Dr Wood, K. S. design the NRL-801 
Unconventional Stellar Aspect Experiment (USA) experiment,
give strategies for using information gathered by X-ray detectors to
determine attitude and position
that use occultation method of traditional celestial navigation,
and timekeeping \citep{w93}.
Dr Hanson, J. E. give the plan of attitude determination algorithm and
timekeeping that use phase-locked loop \citep{h96}. 
Then in the USA Experiment on the ARGOS, 
the X-ray Pulsar-based Navigation System (XPNAVS) 
be first tested which use the Crab pulsar between Feb 1999 and Nov 2000.
In fact, USA experiment is Not Real X-ray pulsar-based navigation experiment
for it is based on occultation method that depend on 
the earth atmospheric models \citep{wdr+02}.
Dr Sheikh, S. I. et al. construct 
the X-ray pulsar-based autonomous navigation theory  
which based on modern spacecraft navigation technique that 
include Kalman filter et al. \citep{s05}.
In the same time, Woodfork, D. W. show that 
the accurate of the position and the clock will be improved 
if using pulsars to aid in a constellation
Signal-In-Space Range Error (SISRE) reduction \citep{w05}.

\begin{figure}
\includegraphics[width=5cm,height=5cm]{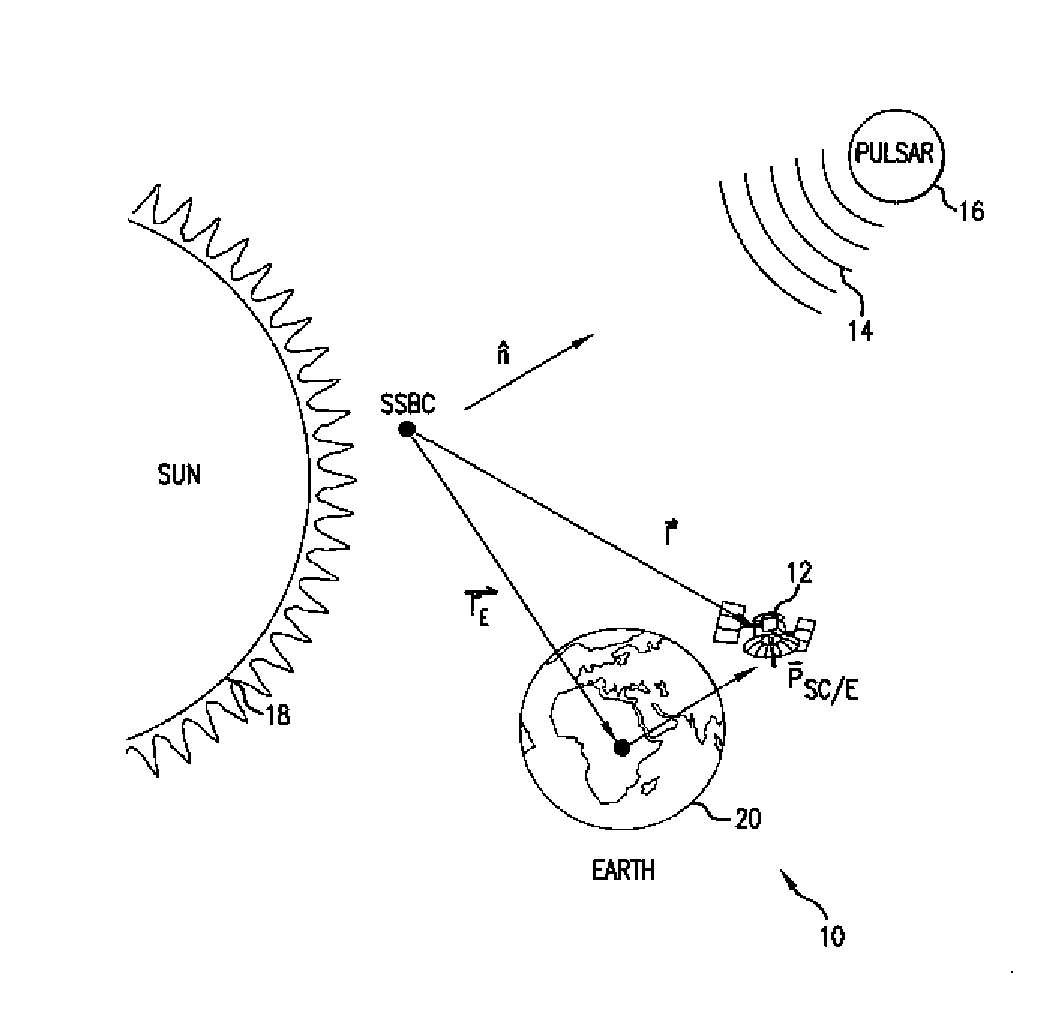}
\vspace{-1.5cm}
\caption[]{}
\label{fig:2D}
\end{figure}

In XPNAVS, the stability of pulsar as one beacon and 
kalman filter to represent the vehicle state
lay the foundation for the navigation.     
Figure. \ref{fig:2D} show the principle of Sheikh's pulsar navigation theory.
Pulsar as the nature lighthouse provide a continuous periodic signal.
Then all signal be normalizing to solar system barycenter coordinates (SSBC).
Though calculate the phase difference of pulsar's times-of-arrival (TOA),
that observed by spacecraft, we will have position and velocity 
by a vector computing in SSBC \citep{sp06, gs07, srw+07, sgp07, suv+04}.
The satellite's amplitude will gain by the same way of star sensor 
\citep{rwp06, spr+06}.

\subsection{Pulsar Clock for Timing }
In 1971, Reichley, et al. described using radio pulsars as clocks \citep{rdm71}.
With researching in-depth, radio astronomer build a stand template 
to pulsar timing \citep{d81, bh86}.
The character of pulsar spin be understood more deeply 
with the timing time increase.
Pulsar especially millisecond pulsars (MSP) be thought 
the nature’s most stable clock \citep{t91}. 
The data show some pulsar stability than atomic clock 
in the timescale greater than one year \citep{mte97}. 
So pulsar time not only is one independent clock to spacecraft \citep{h96} 
but also even can inject to GPS in a long term \citep{w05}.
But currently utilized methods of timing pulse
have errors on the order of hundreds of nanoseconds based upon 
their implementation simplifications, which should be addressed 
if improved accuracies are required \citep{shm07}.

\subsection{Kalman filter for  Position and Velocity }
The kalman filter is an efficient recursive linear filter 
that estimates the state of a dynamic system 
from a series of noisy measurements \citep{k60}.
It can predict the motion of anything for it is recursive,
even the signal have noise, for that use the dynamic state estimate the system. 
(Though Thiele, T. N. and Swerling, P. developed a similar algorithm earlier,
that is Kalman suggest the applicability of his ideas to the problem of 
trajectory estimation for the Apollo program, 
leading to its incorporation in the Apollo navigation computer.)  
Kalman filter is an important topic in control theory 
and control systems engineering, and an important method of
least-squares estimation \citep{s70}.
It is used in a wide range of engineering applications 
which include radar tracking, control system, communication,
guiding and navigation, computer vision, prediction 
in weather and economy, biomedicine, robot et al.
In XPNAVS, that is significant like it in INS and 
the traditional CNS (i.e. star sensor).
We can use navigation kalman filter measure
pulsar range and phase, spacecraft clock,
then compare with the signal which come from pulsar,  
so we will have position and velocity \citep{sp06, gs07, srw+07, sgp07, suv+04}.

In the future XPNAVS, the system noise can not be ignore which origin from
all signal processing \citep{hscg08}.
Pulsar's proper motion must be accurate measurement for 
we must know the beacon position first \citep{mbf+01, ccv+04}.
That must be point which use one pulsar also can 
finish XPNAVS if integrate with INS or star sensor.   

\section{Navigation use Radio Pulsar}
When Downs, G. S. advise that use radio pulsars for 
interplanetary navigation in 1974, 
the antenna and electronic technique can not finish this job 
and pulsar signal process has been understood roughly \citep{lk2005, ls2006}. 
In 1988, Wallace, K. has planned use of radio stars that include pulsar 
for all weather navigation \citep{w88}.
But it is still impossible job on technology in that time.
In fact, just radiometric sextant is widely applied 
on submarine and aircraft carrier et al. in US and Soviet, 
for example the Cod Eye \citep{rs91}.    

Now we think the exist instrument can achieve radio pulsar navigation
although micro-strip antennas can not do it \citep{suv+04}.
The reason that is the technology development and 
pulsar signal process be cognized more deeply \citep{lk2005, ls2006, hr75, r90}.

\subsection{Pulsar signal process in astronomy Vs 
The requires of engineer  project Vs The reliable of technique}
The sensitivity of radio pulsar observation system 
(i.e. the raw limiting flux density) is given by 
the radiometer equation \citep{lk2005, ls2006}:
\begin{equation}
S_{\rm lim} = \frac{\sigma \beta}{(B N_{\rm p} \tau_{\rm obs})^{1/2}}
{\frac{T_{\rm sys}}{G}} ({\frac{W}{P-W}})^{1/2} ~,
\label{eq:s}
\end{equation}
here $\sigma$ is a loss factor, taken to be 1.5 (One-bit sampling
at the Nyquist rate introduces a loss of $\sqrt{2/\pi}$ relative to 
a fully sampled signal. The principal remaining loss results 
from the non-rectangular bandpass of the channel filters.), 
$\beta$ is the detection signal-to-noise ratio threshold, taken to be 5.0, 
$B$ is the receiver bandwidth in Hz, $N_{\rm p}$ is the number of polarizations,
$\tau_{\rm obs}$ is the time per observation in seconds,
$P$ is pulsar period, $W$ is pulse width ($W/P \simeq 0.1$),
$T_{\rm sys}$ is the system temperature, 
$G$ is the telescope gain, $G = A_{\rm e}/(2k_{\rm B})$,
here $A_{\rm e}$ the effective area of a telescope, 
$k_{\rm B}$ is Boltzmann's constant.

From the above described, we can use low-noise receivers, 
a relatively wide bandwidth and long observation times to observed pulsar 
although it is relatively weak radio sources
if there are not a large radio telescope.
Using the equation \ref{eq:s}, 
we use 2 M antenna (If the telescope efficiency is 0.4,
$A_{\rm e} = 0.4 \times \pi(2/2)^2 \simeq 1.256\ {\rm m^2}$, 
$G \simeq 4.55 \times 10^{-4}\ {\rm KJy^{-1}}$),   
$T_{\rm sys}$ is 40 K, 28 GHz bandwidth (2 G - 30 G), $N_{\rm p}$ is 2, 
$\tau_{\rm obs}$ is 36 min,
so we have $S_{\rm lim} \simeq 0.0803\ {\rm Jy} = 80.3\ {\rm mJy}$.
The table  ~\ref{Sources} is the list of the strong radio pulsar source,
it show that we can observed those pulsars 
which use 2 meter antenna in 36 minutes.
If we set $\tau_{\rm obs}$ is 4 min, 4 M antenna, 
we have $S_{\rm lim} \simeq 60.3\ {\rm mJy}$.

The above formula use Jy as unit.
The Jansky (Jy) is a measure of spectral power flux density - 
the amount of RF energy per unit time per unit area per unit bandwidth,
$1\ {\rm Jy} \equiv 10^{-26}\ {\rm W/m^2/Hz}$.
The jansky is not used outside of radio astronomy.
It is not a practical unit for measuring communications signals,
the magnitude is much too small, and is a linear unit,
Very few RF engineers outside of radio astronomy will know what a Jy is.
Because of wide dynamic range encountered the most radio systems,
the power is usually expressed in logarithmic units of 
watts (dBW) or milliwatts (dBm):
${\rm dBW} \equiv 10log_{10}Power_{\rm watts}$,
${\rm dBm} \equiv 10log_{10}Power_{\rm milliwatts}$.
While not comprised of the same units,
we can make some reasonable assumptions to compare a Jy to dBm.
Assumptions bandwidth is 28 GHz (2 G - 30 G), 14 GHz frequency 
($\lambda_{0} = 0.022\ {\rm m}$),
parabolic receive antenna, 
antenna collecting area $ = \pi \times r^2 = 3.14 \times (2/2)^2 
= 3.14\ {\rm m^2}$.
How much is one Jy worth in dBm ?
$P_{mW}= 10^{-26}\ {\rm W/m^2/Hz} \times 28,000,000,000\ {\rm Hz} 
\times 3.14\ {\rm m^2} \times 1000\ {\rm mW/W} 
= 8.82 \times 10^{-13}\ {\rm mW}$,
$P_{\rm dBm} = 10log(8.82 \times 10^{-13}\ {\rm mW}) = -120.5453\ {\rm dBm}$.
Considering the parabolic antenna as a circular aperture gives 
the following approximation for the maximum gain: 
$ G_{\rm dBi} \simeq 10log((9.87 \times D^2)/\lambda_{0}^2$.
in this form, $G$ is power gain over isotropic
$D$ is reflector diameter in same units as wavelength, 
$\lambda_{0} $ is the center of wavelength.
For 2 M diameter and $\lambda_{0} = 0.022\ {\rm m}$, $ G_{\rm dBi} = 49.1153$.
So 1 Jy in 2 M antenna is $-71.43\ {\rm dBm}$.
May be the signal intensity is small for RF engineers,
but for pulsar which like the periodic Gaussian signal, 
we can fold it in the integrate time to increase the pulsar singal sensitivity. 

Pulsar signal suffer dispersion due to the presence of charged particles 
in the interstellar medium. 
The dispersion delay across a bandwidth of $\Delta \nu$ 
centred at a frequency $\nu$ is
\begin{equation}
\label{eq:dm}
\tau_{\rm DM} = 8.30 \times 10^3\, {\rm DM}\, \Delta \nu\, \nu^{-3}\;\;{\rm s},
\end{equation}
where the dispersion measure, DM, is in units of ${\rm cm^{-3}pc}$ 
and the frequencies are in MHz. 
To retain sensitivity, especially for short-period, high-dispersion pulsars, 
the observing bandwidth must be sub-divided into many channels 
to use to incoherent dedispersion 
or achieve the coherent dedispersion \citep{hr75}.
Now a filterbank system have been developed to Digital Filterbank (DFB)
which base field-programmable gate array (FPGA).
The coherent system also enter a new times with 
multi-core and multi-PC cluster development and the price of PC decrease.
In recently, the advantage of graphics processing unit (GPU) and FPGA
in computing be attracted,
if we can fuse Multi-core CPU, GPU and FPGA, 
construct one computing server and use the different advantage of it,
that will easily finish many scientific computation which 
include coherent dedispersion.

In XPNAVS, they use the TOA Measurements of pulsar 
gain the position \citep{sp06}.
In radio waveband, can we not only use the TOA 
but also use the single pulse of pulsar if the antenna is enough big, 
for example, in SIGINT (SIGnals INTelligence) satellite. 
Navigation using pulsar single pulse different with XPNAVS use the TOA,
that will have more precision than use the TOA 
for it direct access to phase information 
that less the error in the measuring process.
Using the above equation \ref{eq:s}, we use 50 M antenna 
(If the telescope efficiency is 0.35, 
$A_{\rm e} = 0.35 \times \pi(50/2)^2 \simeq 686.875\ {\rm m^2}$, 
$G \simeq 2.489 \times 10^{-1}\ {\rm KJy^{-1}}$),   
$T_{\rm sys}$ is 40 K, 28 GHz bandwidth (2 G - 30 G), 
$N_{\rm p}$ is 2, $\tau_{\rm obs}$ is 1 ms,
so we have $S_{\rm lim} \simeq 0.0537\ {\rm Jy} = 53.7\ {\rm mJy}$.
The result show that we can observe the single pulse of pulsar 
in the table ~\ref{Sources} that use 50 M antenna.

In radio pulsar, the strong flux density pulsar usually is young pulsar, 
for example Vela et al., but it take place glitch that is spin faster than past,
that is abnormal phenomenon in pulsar timing model, 
then it become the biggest noise source. 
MSP is a kind of stability pulsar, but it often is weakly.
Many MSP is in binary system, the signal be modulated by orbit effect.
And many MSP in globe cluster, its position unstable for 
the complex gravitational potential.

Navigation of use pulsar just for a continuous pulse signal 
during the mission time which during tens of minutes to several years.
When we penetrate the system of pulsar navigation as one systems engineering,
we think navigation system use radio pulsar is feasible absolutely.
Besides increasing the observation times et al.,
we think some modern digital signal processing (DSP) technique can 
apply to pulsar signal navigation which include 
signal enhancement, signal reconstruct and singularity detection et al.
The hypothesis of pulsar signal is Gaussian 
be used in study pulsar emission geometry 
although it do not be validated directly by observation \citep{wgr+98}. 
In navigation, we just need the information from phase,
so we can magnify the weak pulsar signal through plus a Gaussian signal
or normalizing it to a Gaussian signal on the premise of keep period steadiness.
Glitch can use wavelet to detect in time \citep{dj09}
when use several strong radio pulsar in short time mission.
So we can rule out the interference source, whether using single pulse or TOA.  
The navigation system must leave a copy of raw data to astronomer
for the best filter is construct a good pulsar noise model by it.

\subsection{Pulsar Tracker}
The parabolic dish usually be used in radio astronomy.
But the conventional telescope will bring the control problem
in spacecraft because the big dish is so weight.
Figure. \ref{fig:t} is a radio telescope in Nasu Pulsar Observatory
which the same like Arecibo radio telescope in single dish \citep{tkd+05}.
It will less the difficult of control if use it in vehicle.
So we can use this telescope achieve pulsar tracker like star tracker easily.

\subsection{Pulsar Sensor}
The phased array antenna or radar have seen in recent years breakthroughs 
that lead to capabilities not possible only a few years ago. 
This is exemplified by the development of GaAs integrated microwave
circuits called monolithic microwave integrated circuits (MMIC) 
which makes it possible to build active electronically scanned arrays (AESAs) 
having lighter weight, smaller volume, higher reliability 
and lower cost \citep{b07}. 
Figure. \ref{fig:par} is AESAs of F22 which namely AN/APG-77 and 
built by Northrop Grumman.
This phased array easy achieve pulsar sensor 
(i.e. observe several pulsar simultaneous) 
when it work in passive mode \citep{m97}.
The phased array feed also can apply in pulsar sensor when use one dish.

\subsection{Pulsar Observation in Radio and X-Ray}
When measuring the arrival times of pulsar, the TOA of a fiducial point 
in the rotational phase of a pulsar is the fundamental quantity 
which must be determined. 
This is normally done by comparison with a standard pulse profile $s(t)$. 
The observed pulse profile $p(t)$ can be expressed 
in terms of the standard profile by $p(t) = a + b s(t - \phi) + g(t)$ 
where $a$ is a DC offset, $b$ is a scale factor, 
$\phi$ is a phase shift and $g(t)$ represents noise. 
For the comparison, full use of the available signal to noise 
is most easily achieved by cross-correlating the observed 
and standard profiles in the Fourier domain. 
To do this a very accurate time standard is required 
and is usually obtained from a local hydrogen maser referenced to 
a standard bank of caesium clocks in the ground radio observation \citep{b98}.
But, it is not always possible to have a sufficiently accurate clock 
at the telescope, requiring regular determination of clock offsets.
For example, the absolute time of RXTE’s clock is sufficiently accurate 
to allow this phase of the main X-ray pulse 
to be compared directly with the radio profile \citep{rjl04}.
The reason that is atomic clock so bigger that can not be installed 
in satellite.

A fundamental reason for the provision of contemporary ephemerides 
for timing and searches at other wavelengths (non-radio) is that in many cases, 
less than one photon per pulse period is observed. 
For example the average separation between photons from the Crab 
when observed by EGRET is 10 minutes, 
which corresponds to about 18000 pulse periods \citep{b98}. 
It is the reason that pulsar timing and search in radio waveband in usual.
So pulsar timing in radio have the second advantage 
for have pulsar ephemerides that need long time timing observation.
Pulsar timing in X-ray just about 15 years from RXTE be launched \citep{rjl04}.
Pulsar timing in radio have over 40 years \citep{hlk10}.
If we want to use radio pulsar ephemerides in X-ray,
we must study the phase difference in X-ray and radio
use some years observation \citep{rjl04, lrc+09}.

The frame of reference is the foundation of 
measure the position, orientation, and other properties of objects in it.
Now astronomer have built three reference frame in optic, radio and infrared 
that include FK5 and ICRF2 et al.,
still do not build it in X-ray band \citep{jd99}.
So radio pulsar will direct link to the reference frame in navigation.

\subsection{Integrated Navigation with Pulsar}
Integrated navigation with pulsar in CNS, INS or GNSS,
is realistic path in the future mission.
It will increase the reliability and redundancy of 
navigation or guiding system \citep{zz08}.
The multi-waveband pulsar navigation also is interested, 
for instance, use 1 meter optical telescope \citep{ocg+08} 
or 1.2 meter infrared telescope \citep{rfe+94}
can gain the crab pulsar profile, 
those observation system also easily load in one truck.
In integrated navigation, system analysis and modeling,
system state estimation, filter design, 
information synchronization and system fault tolerance filter design
all is important.
 
\section{Pulsar Navigation in the Solar System}
Like giving different produce in different place by  
navigation systems division of Northrop Grumman Corporation,
pulsar navigation also need use different system 
in each mission \citep{gcsh07, gcs+08, rsg+08}.

\subsection{In the space based and the airborne}
\begin{figure}
\includegraphics[width=5cm,height=8cm]{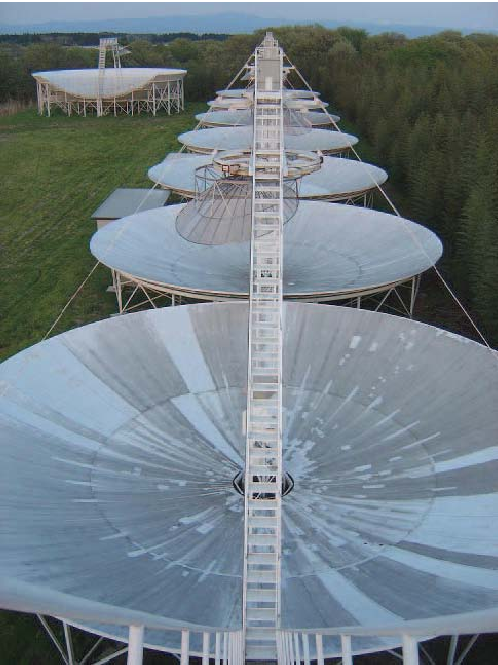}
\vspace{-1.5cm}
\caption[]{}
\label{fig:t}
\end{figure}

\begin{figure}
\includegraphics[width=5cm,height=5cm]{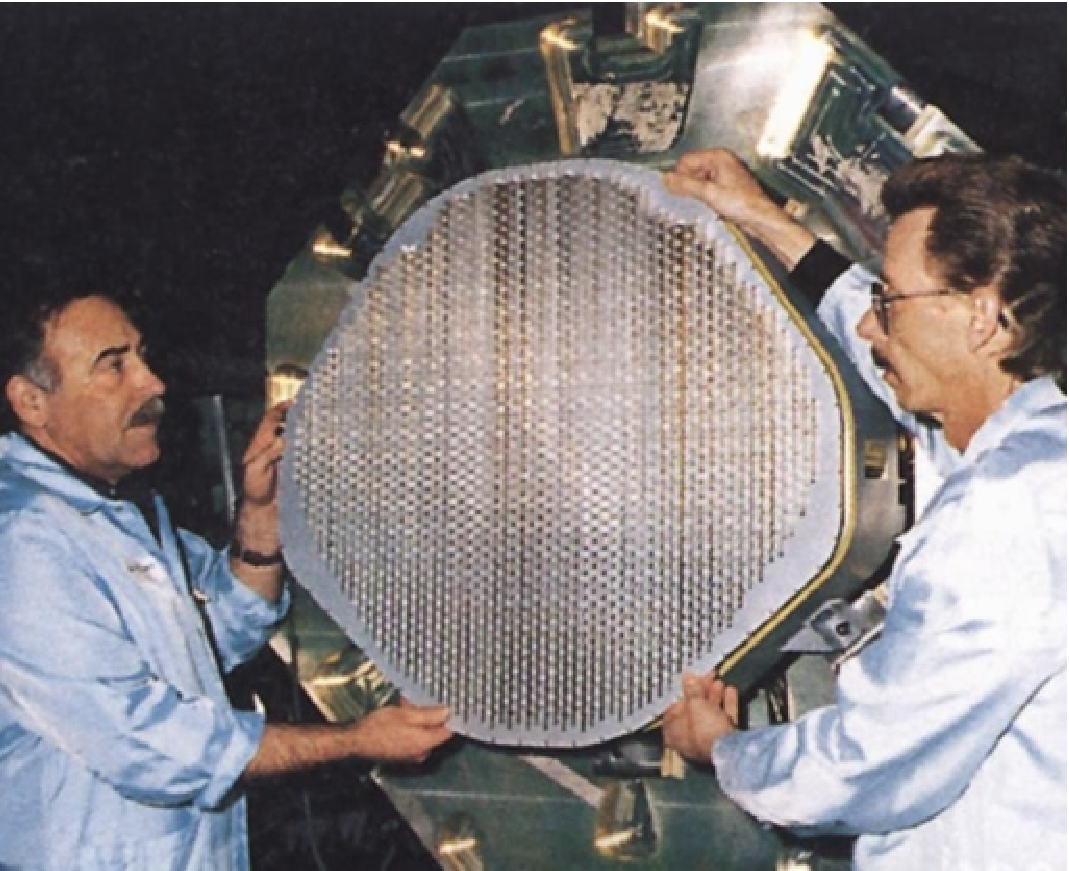}
\vspace{-1.5cm}
\caption[]{}
\label{fig:par}
\end{figure}

In deep space explore, X-ray pulsar tracker suitable for 
most small spacecraft in usually for it even can use a 30 cm detector
have the crab pulsar profile.
But it can not finish pulsar sensor mission now.
We can have spacecraft attitude and position et al. from it.
In International Space Station (ISS) or 
Laser Interferometer Space Antenna (LISA) et al., 
the biggest vehicle, the dish like Figure. \ref{fig:t} is well in mission time.
In radio astronomy, we will have more accurate profile and timing
in the same time which compare with X-ray astronomy 
for the different detection method. 
For some satellites in orbit which include GPS,
the Tracking and Data Relay Satellite System (TDRSS) et al.,
that need precise time, radio tracker like Figure. \ref{fig:t} is a good choice
for it will have a best clock in long term.
SIGINT (SIGnals INTelligence) satellite is the biggest spaceborne antenna
for intelligence-gathering by interception of signals, 
whether between people (i.e., COMINT or communications intelligence) 
or between machines (i.e., ELINT or electronic intelligence), 
or mixtures of the two. Its diameter even has 150 meters.
Radio pulsar signal must be strong noise for it like Crab pulsar to 
the Ballistic Early Warning Site (BMEWS) of US Air Force \citep{s08}.
For some airborne vehicle (F22, J20, B2 et al.) and 
sub-orbit spaceship (X37, X51 et al.),
pulsar sensor like AN/APG-77 of F22 is best choice
for its mission need change attitude frequently.

The Snark (SSM-A-3/B-62/SM-62, Northrop) is the only 
intercontinental surface-to-surface cruise missile (ICCMs) ever deployed 
by the US Air Force, but is operational for only a very short time 
because it was already made obsolete 
by the new Intercontinental Ballistic Missile (ICBMs). 
It first achieve CNS (star tracker) in astronautical.
P-29 (i.e. SS-N-18, Stingray) is submarine-launched ballistic missile
which first achieve one integrated navigation system 
between INS and CNS (star sensor), and it first be launched many simultaneous
for have this integrated navigation system.
The Snack and the Stingray both the large vehicle which 
easily equip pulsar tracker or sensor, 
and it can be used in reconstruct the solar system which include 
dig well in Mars for water et al. \citep{ess07}.
With the distance increase, the radiometric tracking of deep space network (DSN)
will decrease in accuracy \citep{tb05}, 
and it can't work when spacecraft in the other side of sun.
But pulsar can't be effected in that place.
 
\subsection{At the shipboard and submarine}
The big ship all have the radar or antenna for communication et al. \citep{b07}.
Some special ship，for example Yuan Wang tracking ship，
can use to test pulsar navigation in nautical. 
The interesting thing is whether it can be used in submarine
for pulsar can be observed in 12.6 MHz \citep{b87}. 
Project 667 submarines (NATO reporting name Delta) are Soviet-built 
strategic nuclear missile submarines which have 
two VLF/ELF communication buoys. 
Navigation systems include SATNAV, SINS, Cod Eye (radiometric sextant), 
Pert Spring SATCOM \citep{rs91}.
It usually use the enormous antenna net (array underwater) 
to realize VLF/ELF communication,
may be it can receive pulsar signal after a coherent dedispersion.

The higher frequency waves (that is, the shorter the wavelength), 
the more easily be absorbed by water. So radio signal reach deeper than optic, 
but the big antenna net is more difficulty than optic telescope in control.   
In project 667, missile tube about 2 meter, it can observe 
pulsar signal underwater if less one missile and built 
one 2 meter optic or infrared telescope which the same like solar tower.
Those technique carry out will benefit to launch one submarine
to the four gas giants (Jupiter, Saturn, Uranus, and Neptune)
for understand hydro-geology and interiors et al. \citep{ess07}.

\subsection{On the land of planet and lunar}
In the navigation system, the reliability and redundancy is very important.
GPS is a system that spend a lot of money and high maintenance costs.
So Ai GuoXiang et al. develop Chinese Area Positioning System (CAPS) 
that using the communication satellite \citep{ai08}
and Wang AnGuo made the navigation theory that based on 
the measurement of radio signal that 
from celestial and the carrier signal of man-made objects \citep{w07}.
The above system just can application in the earth, 
can not be used in the other planet.
Even if in the earth, the phased array radar can easily moved,
and the 2 meters radio telescope or 1.5 meters optical or infrared
telescope can load in one truck easily.
The same technique can use in lunar rover, Mars rover 
and rover in the others terrestrials, Mercury and Venus.
The virtue is obvious, 
when the rover in the back of the others planet or lunar, 
DSN can not work and 
human can not built GNSS for the other planet in long term. 
So the radio pulsar navigation is one and only method
at any place of the other planet surface day and night in the future explore. 

\section{Pulsar Navigation in the Human evolution to 
the Type II of Kardashev civilizations}
Now Conventional Inertial Reference System (CIRS) is defined in 
the help of the International Celestial Reference Frame (ICRF) that is 
a quasi-inertial reference frame centered at the barycenter of the solar system,
defined by the measured positions of extragalactic sources (mainly quasars).
So it has very high accuracy and reliability.
It will be direct, natural, reliable and accurate, 
if the navigation system be built on the celestial reference system.
Therefor, CNS has some advantages, 
first it is passive measurement in autonomous navigation,
second it has anti-interference ability and is highly reliable,
third it has wide scope of application and the big space of the development,
finally it has simple low-cost equipment 
and facilitate the application and popularization \citep{w07}.

Traditional celestial navigation can be divided into 
optical stars navigation and radio stars navigation. 
In rainy, 
the conventional optical instruments can not observing the celestial bodies, 
the use of navigation time is limited. 
So optical celestial navigation difficult to achieve all-weather work, 
has always a serious obstacle to the application of skills. 
The round-the-clock work is a basic requirement by modern navigation system.
The use of celestial bodies to achieve the astronomical radio navigation can 
be out of adverse weather conditions and restrictions on day and night light. 
As a result, the only way of celestial navigation
is radio technology to accomplish the all-weather navigation. 
Traditional the equipment of radio celestial navigation is radio sextant, 
it only receive a small number of radio signal, 
thus difficult to achieve continuous navigation, 
and just have the low navigation accuracy,  and the equipment size is very big.
So it difficult to application and development.
XNAV is developed by Sheikh et al. in recently,
have achieve the preliminary results in X-ray band. 
Now the European Space Agency (ESA), Russia, France and German
also have begun research it. 
However, these study limited to X-ray band, 
only can be used in spacecraft navigation.
From the above analysis, the small antenna (even two meters) or 
the small optical or infrared telescope (even one meter), 
can receive the stable pulsar signal, 
which means that in radio, optical and infrared bands also can 
achieve the pulsar navigation. 
This work expanded the application range of pulsar navigation,
made it can use in the aerospace, aviation, maritime, ground and underwater. 
So pulsar navigation avoid the disadvantage of 
the traditional radio celestial navigation technology.

In recently, Coll, B. and Tarantola, A. give the analysis of pulsar navigation 
in the Milky Way that base general relativity theory \citep{ct09}.
If we can understand the effect of pulsar emission area, 
may be we can use it navigation in the Milky way.   

Soviet astronomer Kardashev, N. S. propose 
a scheme for classifying advanced technological civilizations.
He identified three possible types and distinguished 
between them in terms of the power 
they could muster for the purposes of interstellar communications. 
A Type I civilization would be able to marshal energy resources 
for communications on a planet-wide scale, 
equivalent to the entire present power consumption of the human race,
or about $10^{16}$ watts. 
A Type II civilization would surpass this by 
a factor of approximately ten billion, making available $10^{26}$ watts, 
by exploiting the total energy output of its central star. 
Finally, a Type III civilization would have evolved far enough to 
tap the energy resources of an entire galaxy. 
This would give a further increase by at least a factor of 10 billion 
to about $10^{36}$ watts \citep{k64}. 
Carl Sagan estimate that, on this more discriminating scale, 
the human race would presently qualify as roughly a Type 0.7. 
In the Age of Discovery, that is CNS make human freely voyage in the sea.
So it make human civilization increase to higher type. 

Now the Second Age of Discovery or Age of Exploration 
in the solar system is beginning,
pulsar as nature beacon in the Milky Way will make human freely fly 
in the space of solar system. 
Recently, the first Falcon 9 flight is successfully launched
on June 4, 2010 with a successful orbital insertion.
It is a spaceflight launch system that uses rocket engines designed 
and manufactured by SpaceX company.
Many private company of the aerospace industry
for instance SpaceX and SpaceDev et al.,
need one low cost and reliable navigation system.
Pulsar navigation give a path which do not depend on DSN,
so it less huge cost in the outer space and the interplanetary navigation.
It make the spacecraft of the private company
not only enter the outer space but also voyage to the other planet.
After some pioneer explore, if we can find one mode to gain profit, 
may be tour or dig ore that include ${\rm He_3}$ in lunar and 
diamonds in Uranus and Neptune \citep{kdd08}, the new manufacturing 
about space travel will lead people into a new economic era,
and the real Second Age of Discovery or Age of Exploration will begun.
That is extraordinary in the human evolution to 
type II of Kardashev civilizations.

\section*{Acknowledgements}
The author thank DARPA make someone invent Internet and open it for public.

\bibliographystyle{apsr}
\bibliography{IEEEfull,pnbib}

\begin{table*}[htbp]
\caption{The strong sources of radio pulsar. \label{Sources}}
\begin{center}
\begin{tabular}{l|c|c|c|c|c}
\hline
Name       & Period & DM               & W50     & S400/S1400 &relation with \\
Pulsar     & (S)    &(${\rm cm^{-3}pc}$) & (ms)   & (mJy)      &Glich(G)\\
\hline
J0332+5434    &0.71452  &26.833 &6.6 &1500/203 &                \\   
J0953+0755    &0.25306 &2.958  &9.5 &400/84   &                \\
J0747$-$4715  &0.00576 &2.64476 &0.969 &550/142 &        \\
J0738$-$4042  &0.37492  &160.8  &29   &190/80   &                  \\
J0835$-$4510  &0.08933  &67.99  &2.1  &5000/100 & G \\
J1456$-$6843  &0.26338 &8.6    &12.5 &350/80   &  G \\
J1644$-$4559  &0.45506  &478.8  &8.2  &375/310  & G     \\
\hline
\end{tabular}
\end{center}
The strong source in radio pulsar, all can use to navigation 
when aovid glitch noise. $1\ {\rm Jy} \equiv 10^{-26}\ {\rm W/m^2/Hz}$.  
\end{table*}

\noindent {\bf Fig. 1.} A schematic view of the subject system for navigation
  utilizing sources of pulsed celestial radiation \citep{spw+07}.

\noindent {\bf Fig. 2.} Photograph of the array at Nasu Radio Interferometer. 
Eight equally spaced, 20 m diameter, fixed spherical antennas are shown in this figure.

\noindent {\bf Fig. 3.} The AN/APG-77 is a multi-function radar installed on 
the F-22 Raptor fighter aircraft. The radar is built by Northrop Grumman.
The figure come from wiki: \url{http://en.wikipedia.org/wiki/AN/APG-77}, accessed Nov 30 2008

\end{document}